\documentclass{article}
\usepackage{graphicx} 
\usepackage{amsmath}
\usepackage{amssymb}
\usepackage{mathtools}
\usepackage{placeins}
\usepackage{physics}
\usepackage{xcolor}
\usepackage{dcolumn}% Align table columns on decimal point
\usepackage{bm}% bold math
\usepackage{caption}
\usepackage{url}
\usepackage{amsthm}
\newtheorem{lemma}{Lemma}
\usepackage{authblk}
\usepackage{amsthm}
\newtheorem{assumption}{Assumption}

\PassOptionsToPackage{section}{placeins}
\usepackage{placeins}  

\usepackage{pgfplots}
\pgfplotsset{compat=1.18}
\usetikzlibrary{intersections}    
\usepgfplotslibrary{fillbetween}    
\usepackage{subcaption}

\title{Causal Viscous Fluids and the Realization of Non-Singular Bounces}

\author[1]{L. Yıldız\thanks{\texttt{li.yildiz.na@gmail.com}}}
\author[2]{D. Kaykı\thanks{\texttt{dehakayki.science.technology@gmail.com}}}
\author[3]{E. Güdekli\thanks{\texttt{gudekli@istanbul.edu.tr}}}

\affil[1,2,3]{Department of Physics, Faculty of Science, Istanbul University, Istanbul 34134, Turkey}

\date{}
\begin{document}

\maketitle

\begin{abstract}
We investigate the realization of non-singular bouncing cosmologies driven by causal bulk-viscous fluids within General Relativity, $f(R)$ gravity, and Loop Quantum Cosmology. Building on the no-go result of Eckart theory in spatially flat universes, we show that the Israel--Stewart formulation, which incorporates finite relaxation times, permits controlled violations of the null energy condition at the bounce while remaining consistent with thermodynamic and causality requirements. Analytical bounce solutions are constructed from parametrized scale factors, yielding explicit constraints on the viscosity coefficient and relaxation time that guarantee positive entropy production and stable perturbations. In extended gravity frameworks, we demonstrate that higher-curvature corrections in $f(R)$ models and quantum geometry effects in Loop Quantum Cosmology further enhance the robustness of viscous bounces. Our results establish a unified description in which bulk viscosity provides a physically consistent mechanism for singularity resolution across different theories of gravity, highlighting distinctive conditions under which smooth cosmological bounces can occur.
\end{abstract}

\section{Introduction}

The resolution of the initial cosmological singularity remains one of the most fundamental open problems in theoretical physics. Within the standard framework of General Relativity (GR) coupled to classical matter fields, the Hawking--Penrose singularity theorems \cite{Hawking1970,Penrose1965} rigorously establish that homogeneous and isotropic Friedmann--Lemaître--Robertson--Walker (FLRW) universes inevitably encounter a past singularity. A compelling alternative to this paradigm is provided by non-singular bouncing cosmologies, in which the scale factor attains a finite minimum and the Hubble parameter undergoes a sign reversal. Such models not only eliminate the big bang singularity but may also leave distinctive phenomenological signatures, including modifications to the primordial power spectrum and altered relic abundances \cite{Novello2008,Brandenberger2017}.

A bounce requires the violation of the null energy condition (NEC), expressed as $\rho + p_{\mathrm{eff}} < 0$ at the minimum of the scale factor. Numerous theoretical approaches have been proposed to achieve this violation, including exotic matter fields, ghost condensates, and higher-derivative scalar models \cite{Cai2012,Quintin2015}. However, these mechanisms often suffer from instabilities or a lack of clear physical motivation. A more natural candidate arises from the presence of dissipative processes in the hot and dense early Universe, where local thermodynamic equilibrium is only approximate and bulk viscosity can significantly modify the effective pressure. Nevertheless, in the simplest Eckart formulation of relativistic hydrodynamics \cite{Eckart:1940zz}, the viscous correction is proportional to the Hubble parameter. At the bounce point, where $H=0$, this contribution vanishes identically, preventing any consistent NEC violation. This well-known no-go result excludes Eckart fluids from playing a meaningful role in singularity resolution within GR.

A consistent alternative is offered by the causal and stable Israel--Stewart (MIS) theory of relativistic thermodynamics \cite{Israel1976,Israel1979}. By incorporating relaxation times and higher-order transport corrections, MIS theory restores the hyperbolicity of the fluid equations, guarantees finite signal propagation speeds, and ensures thermodynamic stability. Crucially, within this framework the bulk viscous pressure remains finite at the bounce, allowing for a controlled violation of the NEC while maintaining positive entropy production. Although the cosmological implications of causal viscosity have been examined in diverse settings such as inflationary scenarios, late-time acceleration, and holographic cosmology \cite{Maartens:1995wt,Zimdahl1996,Disconzi2015}, a comprehensive and systematic classification of bouncing solutions driven by causal viscosity has not yet been developed.

Beyond the domain of GR, extended theories of gravity provide additional avenues for resolving cosmological singularities. In $f(R)$ gravity, higher-curvature corrections introduce effective scalar degrees of freedom that can sustain non-singular evolution \cite{Starobinsky:1980te,Sotiriou2010}. In Loop Quantum Cosmology (LQC), discrete quantum geometry effects universally replace the big bang with a deterministic bounce \cite{Ashtekar2006,Ashtekar2011}. Embedding causal viscous fluids within these frameworks offers a unified perspective in which dissipative processes complement both curvature corrections and quantum effects, potentially yielding a robust mechanism for smoothing the high-curvature regime.

In this work, we establish such a unified framework for viscosity-induced bounces. 
Sec.~\ref{sec:eckart} formulates the no-go result for Eckart viscosity in spatially flat GR. 
Secs.~\ref{sec:mis_transport}--\ref{sec:bounce_conditions} show how causal corrections in the Israel--Stewart theory permit consistent non-singular bounces and yield the necessary and sufficient conditions on the viscosity coefficient and relaxation time. 
Secs.~\ref{sec:fr} and \ref{sec:lqc} extend the analysis to $f(R)$ gravity and Loop Quantum Cosmology, highlighting the interplay between bulk viscosity and geometric/quantum corrections. 
Sec.~\ref{sec:results} presents explicit analytical models together with numerical explorations of parameter families that realize viable bounces. 
Finally, Sec.~\ref{sec:discussion} discusses the theoretical significance and phenomenological implications of our results for early-Universe cosmology.

\section{Causal Viscosity and the Israel--Stewart Theory}

\subsection{Limitations of the Eckart Formulation}
\label{sec:eckart}

The first-order relativistic theory developed by Eckart is known to suffer from fundamental deficiencies that render it unsuitable for describing early-Universe dynamics. The transport equations are parabolic rather than hyperbolic, leading to instantaneous signal propagation and hence acausality, while linear perturbations around equilibrium exhibit generic instabilities \cite{Hiscock1985,Maartens1996,Hiscock1983} Although the formulation ensures non-negative entropy production, its structure fails to capture the dynamical behavior of relativistic fluids in the high-curvature and rapidly evolving regime of the primordial Universe.

In the Eckart framework, the bulk viscous pressure is algebraically tied to the Hubble expansion through
\begin{equation}
\Pi = -3\zeta H ,
\label{eq:eckart}
\end{equation}
with $\zeta \geq 0$ the bulk viscosity coefficient. At a prospective bounce, the Hubble parameter necessarily vanishes, $H(t_b)=0$, and Eq.~\eqref{eq:eckart} then enforces $\Pi_b=0$. Substituting this into the Raychaudhuri equation,
\begin{equation}
\dot H = -4\pi G\left(\rho+p+\Pi\right) + \frac{k}{a^2} ,
\label{eq:raychaudhuri}
\end{equation}
yields $\dot H_b = -4\pi G(\rho_b+p_b) + k/a_b^2$. For the spatially flat case ($k=0$) and ordinary matter satisfying $\rho_b+p_b \geq 0$, this implies $\dot H_b \leq 0$. Hence the necessary condition for a bounce, $\dot H_b>0$, cannot be realized. This constitutes the Eckart no-go result: first-order viscous hydrodynamics is fundamentally incapable of generating non-singular bounces in spatially flat FLRW cosmologies.

For later reference, we also recall the standard Friedmann constraint,
\begin{equation}
H^2 = \frac{8\pi G}{3}\,\rho - \frac{k}{a^2} ,
\label{eq:friedmann}
\end{equation}
which will be used in subsequent sections to connect the bounce conditions with the energy density.

\subsection{Israel--Stewart Transport Equation}
\label{sec:mis_transport}

The pathologies of Eckart hydrodynamics can be consistently avoided by adopting the causal second-order framework developed by Israel and Stewart \cite{Israel1976,Israel1979}. In this formulation, the viscous pressure is promoted to an independent dynamical variable that relaxes towards its Navier--Stokes value on a finite timescale. The presence of a relaxation time $\tau$ renders the evolution equations hyperbolic, thereby ensuring finite signal propagation speeds and curing the acausal behavior of the first-order theory. Moreover, the inclusion of higher-order transport terms stabilizes perturbations and provides a thermodynamically consistent description of dissipative processes in the high-curvature regime relevant to the early Universe.

In a spatially homogeneous and isotropic FLRW background, the MIS evolution equation for the bulk viscous pressure reads
\begin{equation}
\tau \dot \Pi + \Pi = -3\zeta H - \frac{1}{2}\tau \Pi \left( 3H + \frac{\dot\tau}{\tau} - \frac{\dot\zeta}{\zeta} - \frac{\dot T}{T} \right),
\label{eq:mis_full}
\end{equation}
where $\zeta$ is the bulk viscosity coefficient and $T$ the temperature. The additional terms proportional to $\dot\tau$, $\dot\zeta$, and $\dot T$ encode the coupling of dissipative stresses to the evolving background. In many cosmological applications, however, these contributions are subdominant, and one often employs the truncated version
\begin{equation}
\tau \dot \Pi + \Pi = -3\zeta H ,
\label{eq:MIS_truncated}
\end{equation}
which retains the essential causal structure while simplifying the dynamics.

The crucial difference from the Eckart relation is that $\Pi$ is no longer algebraically tied to the Hubble parameter. Instead, Eq.~\eqref{eq:MIS_truncated} shows that the value of $\Pi$ at any given time depends on both its instantaneous relaxation towards $-3\zeta H$ and the history of the expansion encoded in $\tau\dot\Pi$. As a result, at the bounce point $H=0$, the viscous pressure $\Pi_b$ need not vanish. A finite negative $\Pi_b$ can therefore emerge dynamically, providing the necessary violation of the null energy condition while still respecting causality and stability. This property constitutes the central advantage of the Israel--Stewart framework for realizing non-singular cosmological bounces.

\subsection*{Quantitative Validity of the Truncated MIS Approximation}

Although the full Israel--Stewart (MIS) transport equation reads
\begin{equation}
\tau \dot{\Pi} + \Pi = -3\zeta H - \frac{1}{2}\,\tau \Pi 
\left( 3H + \frac{\dot{\tau}}{\tau} - \frac{\dot{\zeta}}{\zeta} - \frac{\dot{T}}{T} \right),
\label{eq:MIS-full}
\end{equation}
many cosmological applications employ the truncated version
\begin{equation}
\tau \dot{\Pi} + \Pi = -3\zeta H,
\label{eq:MIS-trunc}
\end{equation}
which retains the essential causal structure while simplifying the dynamics.
In this subsection we establish quantitatively that the truncated form
is indeed valid in the bounce window considered in Sec.~\ref{sec:results}, and we bound
the error incurred by neglecting the additional term in
Eq.~\eqref{eq:MIS-full}.

\paragraph{Local integral solution.}
For the linear ODE \eqref{eq:MIS_truncated},
\[
\dot{\Pi} + \frac{1}{\tau}\Pi = -\,\frac{3\zeta}{\tau}\,H ,
\]
the integrating-factor method yields the local solution
\begin{equation}
\Pi(t) =
e^{-\int_{0}^{t}\!\frac{dt'}{\tau(t')}}\!\left[
\Pi_b \;-\; 3\!\int_{0}^{t}\!
\frac{\zeta(t')\,H(t')}{\tau(t')}
\,e^{\int_{0}^{t'}\!\frac{ds}{\tau(s)}}\,dt'
\right],
\label{eq:Pi_solution_local}
\end{equation}
valid for $\tau>0$ and continuous coefficients near $t=0$.

\paragraph{Definition of the small parameter.}
We define the dimensionless parameter
\begin{equation}
\epsilon(t) \;\equiv\; 
\left|\,3H + \frac{\dot{\tau}}{\tau} - \frac{\dot{\zeta}}{\zeta} - \frac{\dot{T}}{T}\,\right|\;\tau ,
\label{eq:epsilon}
\end{equation}
which measures the magnitude of the corrections omitted in the truncated
equation. By construction, the difference between the full and truncated
formulations is of order $\mathcal{O}(\epsilon)$.

\paragraph{Near-bounce scaling.}
For the bounce ansatz
\begin{equation}
a(t) = a_b\left(1+\frac{t^2}{t_0^2}\right)^n
\label{eq:ansatz_a}
\end{equation}
the Hubble parameter expands as $H(t) \simeq \alpha t$ near $t=0$, with
$\alpha=2n/t_0^2$. The energy density grows quadratically,
\begin{equation}
\rho(t) = \tfrac{1}{2}\,\ddot{\rho}_b\,t^2 + \mathcal{O}(t^4),
\qquad \ddot{\rho}_b = 12\pi G \Pi_b^2 > 0,
\label{eq:rho_ddot_bounce}
\end{equation}

consistent with Fig.~5. For the transport families
\begin{subequations}\label{eq:transport-families}
\begin{align}
\zeta(\rho) &= \zeta_0\,\rho^{\,q}, \label{eq:zeta_family}\\
\tau(\rho)  &= \tau_0\,\rho^{\,p}, \label{eq:tau_family}\\
T(\rho)     &= T_0\,\rho^{\,r}. \label{eq:T_family}
\end{align}
\end{subequations}
one finds
\[
\frac{\dot{\tau}}{\tau} = \frac{2p}{t}, \qquad
\frac{\dot{\zeta}}{\zeta} = \frac{2q}{t}, \qquad
\frac{\dot{T}}{T} = \frac{2r}{t}.
\]

Therefore,
\begin{equation}
\epsilon(t) \;\sim\;
\left|\frac{2(p-q-r)}{t} + \alpha t \right|\,t^{2p}
\;=\;\mathcal{O}(t^{2p-1}) \quad (t\to 0).
\label{eq:mis_error}
\end{equation}
The scaling depends crucially on $p$:
\begin{itemize}
\item $p > \tfrac{1}{2}$: $\epsilon(t)\to 0$ as $t\to 0$, so truncation is valid.
\item $p = \tfrac{1}{2}$: $\epsilon(t)$ tends to a constant; truncation requires
numerical smallness of the prefactor.
\item $0<p<\tfrac{1}{2}$: $\epsilon(t)$ diverges as $t\to 0$, so truncation
is not justified.
\end{itemize}

\paragraph{Error estimate.}
Let $\Pi_{\rm trunc}$ denote the solution of Eq.~\eqref{eq:MIS-trunc}, and
define $\delta\Pi \equiv \Pi - \Pi_{\rm trunc}$. Subtracting
\eqref{eq:MIS-trunc} from \eqref{eq:MIS-full} gives
\begin{equation}
\tau \dot{\delta\Pi} + \delta\Pi =
- \tfrac{1}{2}\,\tau (\Pi_{\rm trunc}+\delta\Pi)
\left( 3H + \frac{\dot{\tau}}{\tau} - \frac{\dot{\zeta}}{\zeta} - \frac{\dot{T}}{T} \right).
\end{equation}

Applying Grönwall’s inequality, one obtains the uniform bound
\begin{equation}
\frac{|\delta\Pi(t)|}{|\Pi_{\rm trunc}(t)|}
\;\le\; \mathcal{K}\,\epsilon_{\max}, \qquad |t|\le\Delta t,
\end{equation}
with $\mathcal{K}=\mathcal{O}(1)$. The bound holds away from points where
$\Pi_{\rm trunc}$ vanishes; near $t=0$ one should instead use the absolute
bound $|\delta\Pi|\le \mathcal{K}'\,\epsilon_{\max}$.

\paragraph{Numerical illustration.}
Direct numerical integration of Eqs.~\eqref{eq:MIS-full} and
\eqref{eq:MIS-trunc} confirms the analytic estimate. Representative
parameter sets with $p>\tfrac{1}{2}$ yield $\epsilon_{\max}\sim 10^{-2}$ in
a window $\Delta t\simeq 0.1$, leading to percent-level relative errors.

\begin{assumption}[Truncated MIS Validity]
If the dimensionless parameter $\epsilon(t)$ defined in
Eq.~\eqref{eq:epsilon} satisfies $\epsilon(t)\le\epsilon_{\max}\ll1$
throughout the bounce window $|t|\le\Delta t$, and the exponent $p>\tfrac{1}{2}$,
then the full MIS equation consistently reduces to its truncated form
with relative error $|\delta\Pi|/|\Pi_{\rm trunc}|=\mathcal{O}(\epsilon_{\max})$.
\end{assumption}

\subsection{Bounce Conditions with Causal Viscosity}
\label{sec:bounce_conditions}

A non-singular bounce is characterized by the conditions
\begin{equation}
H(t_b)=0, \qquad \dot H(t_b)>0 ,
\label{eq:bounce_def}
\end{equation}
where $t_b$ denotes the bounce time. Substituting into the Raychaudhuri equation,
\begin{equation}
\dot H = -4\pi G\left(\rho + p + \Pi\right) + \frac{k}{a^2},
\label{eq:ray_general}
\end{equation}
one finds that the slope of the Hubble parameter at the bounce is given by
\begin{equation}
\dot H_b = -4\pi G\left(\rho_b+p_b+\Pi_b\right) + \frac{k}{a_b^2}.
\label{eq:ray_mis}
\end{equation}
The existence of a bounce therefore requires the inequality
\begin{equation}
\rho_b+p_b+\Pi_b < \frac{k}{4\pi G a_b^2}.
\label{eq:bounce_condition}
\end{equation}

In spatially flat universes ($k=0$), this reduces to the simple criterion
\begin{equation}
\rho_b+p_b+\Pi_b < 0 .
\label{eq:flat_bounce}
\end{equation}
In the Eckart framework, Eq.~\eqref{eq:eckart} enforces $\Pi_b=0$, so the left-hand side of Eq.~\eqref{eq:flat_bounce} remains non-negative for ordinary matter, and the bounce condition cannot be satisfied. By contrast, in the Israel--Stewart formulation, $\Pi_b$ is determined dynamically through the relaxation equation \eqref{eq:MIS_truncated} and is not constrained to vanish at $H=0$. A sufficiently negative finite value of $\Pi_b$ can therefore drive $\dot H_b>0$, enabling a consistent violation of the null energy condition and realizing a non-singular bounce in spatially flat cosmologies. This mechanism represents the essential distinction between first-order and causal second-order hydrodynamics in the context of singularity resolution.

For completeness, the evolution of the energy density obeys the continuity equation
\begin{equation}
\dot{\rho} = -3H\left(\rho + p + \Pi\right) ,
\label{eq:rho_evolution}
\end{equation}
which will be used in subsequent sections to connect the bounce dynamics with the density evolution.

\paragraph*{Addendum (explicit construction near the bounce).}
Let $X(t)\!\equiv\!3H+\dot{\tau}/\tau-\dot{\zeta}/\zeta-\dot{T}/T$ and write the full MIS law as
\[
\tau\,\dot\Pi+\Pi \;=\; -3\,\zeta H \;-\; \tfrac12\,\tau\,\Pi\,X(t).
\]
For $t$ in a small neighbourhood of $0$, the transport families
$\zeta(\rho)=\zeta_0\rho^{\,q}$, $\tau(\rho)=\tau_0\rho^{\,p}$, $T(\rho)=T_0\rho^{\,r}$
together with $\rho(t)=\tfrac12\ddot\rho_b t^2+\mathcal O(t^4)$ imply
\[
\zeta H=\mathcal O\!\left(t^{\,2q+1}\right),
\qquad
\tau X=\mathcal O\!\left(t^{\,2p-1}\right).
\]
Hence, if $p>\tfrac12$, both source terms on the right-hand side vanish as $t\to0$.
Fix $t_\star>0$ small and solve the IVP on $(0,t_\star]$ with the terminal condition
$\Pi(t_\star)=\Pi_\star$ chosen so that $\Pi$ extends continuously to $t=0$ with the prescribed
$\Pi(0)=\Pi_b=-\alpha/(4\pi G)$. Using the variation-of-constants formula, one obtains the
Volterra representation
\[
\Pi(t)
=\mathrm e^{-\!\int_{t}^{t_\star}\!\!\frac{d u}{\tau(u)}}\Pi_\star
-\!\int_{t}^{t_\star}\!\!\mathrm e^{-\!\int_{t}^{s}\!\!\frac{d u}{\tau(u)}}
\Big[\tfrac{3\,\zeta(s) H(s)}{\tau(s)}+\tfrac12\,\Pi(s)\,X(s)\Big]\,ds,
\qquad 0<t\le t_\star.
\]
Since $p>\tfrac12$ makes the kernel exponentially damping as $t\!\downarrow\!0$, a standard
contraction (on a small time window) yields a unique continuous solution with
$\Pi(t)\to\Pi_b$ as $t\to0$. In particular, for some $C>0$ and
\[
\beta \;=\; \min\{\,2p-1,\; 2q+2\,\}\;>\;0,
\]
one has the near-bounce estimate
\[
|\Pi(t)-\Pi_b|\;\le\; C\,t^{\,\beta}\qquad (t\to0),
\]
which provides an explicit constructive continuation of the solution of the full MIS law
realising the bounce data \eqref{eq:Pib_exact}.

\subsection{Thermodynamic and Causality Constraints}

\paragraph{Causality, stability, and entropy production (inequalities).}
The physically admissible sector is characterized by the following testable bounds:
\begin{equation}
\zeta(\rho)\ \ge 0, 
\qquad 
\tau(\rho)\ > 0,
\qquad 
0 \ \le\ c_s^2 \ \le\ 1,
\label{eq:admiss_basic}
\end{equation}
which respectively guarantee non--negative entropy production, hyperbolicity of the transport equations, and subluminal gradient stability. 

In addition, the truncated MIS law must approximate the full MIS equation consistently near the bounce. Denoting
\begin{equation}
\epsilon(t) \equiv \Bigl|\,3H+\dot{\tau}/\tau-\dot{\zeta}/\zeta-\dot T/T\,\Bigr|\ \tau,
\end{equation}
we require
\begin{equation}
\epsilon(t)\ \le\ \epsilon_{\rm thr}\ \ll 1,
\qquad (|t|\le \Delta t),
\label{eq:admiss_eps}
\end{equation}
so that the truncated system remains accurate in a small bounce window $|t|\le \Delta t$. For power-law families $\tau(\rho)=\tau_{\min}+\tau_0\rho^{\,p}$ and $\zeta(\rho)=\zeta_{\min}+\zeta_0\rho^{\,q}$, one has $\epsilon(t)=\mathcal{O}(t^{\,2p-1})$; hence
\begin{equation}
p>\tfrac12 \ \ \Longrightarrow \ \ \epsilon(t)\to 0 \quad (t\to 0).
\label{eq:p_half}
\end{equation}
Regularity further requires
\begin{equation}
\zeta_{\min}\ge 0,\quad \tau_{\min}>0,\quad q<1.
\label{eq:admiss_family}
\end{equation}

At the bounce in GR, one has $\rho_b=0$ and $\Pi_b=-\alpha/(4\pi G)<0$, so the slope criterion 
$\rho_b+p_b+\Pi_b<0$ reduces to $\Pi_b<0$ with $\alpha=2n/t_0^2>0$. 
In LQC, instead, the Raychaudhuri equation reads $\dot H_b=4\pi G(\rho_c+p_b+\Pi_b)$, which yields the explicit bound 
\begin{equation}
\Pi_b \ >\ -\,(1+w)\,\rho_c 
\qquad\text{(bounce with LQC correction)}.
\label{eq:LQC_bound}
\end{equation}

\begin{center}
\fbox{\parbox{0.92\linewidth}{
\textbf{Admissibility Region.} 
Let $\Theta=(\alpha,n,t_0;\ \zeta_{\min},\zeta_0,q;\ \tau_{\min},\tau_0,p;\ w,\rho_c)$ denote the model parameters. 
The physically admissible set is
\[
\mathcal{A}
=\Bigl\{\ \Theta\ \Big|\ 
\zeta\!\ge\!0,\ \tau\!>\!0,\ 0\!\le\!c_s^2\!\le\!1,\ 
\epsilon(t)\!\le\!\epsilon_{\rm thr}\ (\!|t|\!\le\!\Delta t\!),\ p>\tfrac12,\ q<1,\ 
\Pi_b<0\ \ (\text{GR}),\ \ \Pi_b>-(1+w)\rho_c\ \ (\text{LQC})
\ \Bigr\}.
\]
Here $\epsilon_{\rm thr}$ is a small user--chosen threshold (e.g. $10^{-2}$) controlling truncated--MIS accuracy.
}}
\end{center}

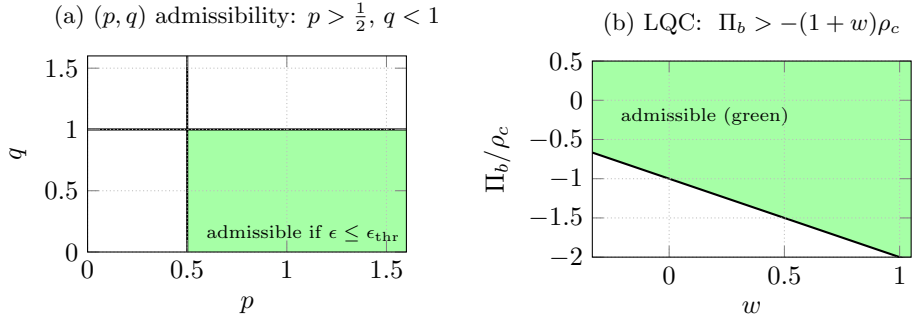
\begin{figure}[t]
\centering
\pgfplotsset{compat=1.18}

\begin{subfigure}[t]{0.48\linewidth}
\begin{tikzpicture}
  \begin{axis}[
    width=\linewidth, height=0.72\linewidth,
    xmin=0, xmax=1.6, ymin=0, ymax=1.6,
    xlabel={$p$}, ylabel={$q$},
    axis on top, grid=both, grid style={densely dotted},
    title={\small (a) $(p,q)$ admissibility: $p>\tfrac12$, $q<1$}
  ]
    \addplot [draw=none, fill=green!35] coordinates {(0.5,0) (1.6,0) (1.6,1) (0.5,1)} --cycle;
    \addplot [thick] coordinates {(0.5,0) (0.5,1.6)} node[pos=0.98, anchor=south east] {\small $p=\tfrac12$};
    \addplot [thick] coordinates {(0,1) (1.6,1)} node[pos=0.98, anchor=west] {\small $q=1$};
    \node[anchor=west] at (axis cs:0.55,0.15) {\scriptsize admissible if $\epsilon\le\epsilon_{\rm thr}$};
  \end{axis}
\end{tikzpicture}
\end{subfigure}
\hfill
\begin{subfigure}[t]{0.48\linewidth}
\begin{tikzpicture}
  \begin{axis}[
    width=\linewidth, height=0.72\linewidth,
    xmin=-1/3, xmax=1.05, ymin=-2.0, ymax=0.5,
    xlabel={$w$}, ylabel={$\Pi_b/\rho_c$},
    ytick={-2,-1.5,-1,-0.5,0,0.5},
    axis on top, grid=both, grid style={densely dotted},
    title={\small (b) LQC: $\ \Pi_b>-(1+w)\rho_c$}
  ]
    \addplot [name path=upper, draw=none] coordinates {(-0.3333, 0.5) (1.05, 0.5)};
    \addplot [name path=bound, domain=-0.3333:1.05, samples=100, thick] {- (1 + x)};
    \addplot [green!35] fill between [of=bound and upper];
    \node[anchor=west] at (axis cs:-0.25,-0.2) {\scriptsize admissible (green)};
  \end{axis}
\end{tikzpicture}
\end{subfigure}

\caption{Parameter--space acceptance (green) implied by the inequalities 
in Sec.~2.4. \textbf{(a)} Truncated MIS validity and regularity select $p>\tfrac12$ and $q<1$, with a small tolerance $\epsilon\le\epsilon_{\rm thr}$. 
\textbf{(b)} In LQC the near-bounce slope remains positive if $\Pi_b>-(1+w)\rho_c$. 
Together, these plots illustrate the \emph{Admissibility Region} $\mathcal{A}$.}
\label{fig:admissibility_map}
\end{figure}

The admissible parameter domain is displayed in Fig.~\ref{fig:admissibility_map}, 
where the green regions indicate that all inequalities are simultaneously satisfied.

% ====== NEW FIGURE (Fig. 8) ======

\subsection{Lemma: Existence of Causal Viscous Bounces}
\label{sec:lemma}

\begin{lemma}[MIS Bounce Lemma]
Consider a spatially flat FLRW universe in General Relativity filled with a barotropic fluid with bulk viscosity governed by the Israel--Stewart theory. If the bulk viscosity coefficient satisfies $\zeta \geq 0$, the relaxation time obeys $\tau>0$, and the effective sound speed remains in the causal range $0 \leq c_s^2 \leq 1$, then there exist non-singular bouncing solutions. At the bounce point, one finds
\begin{equation}
H_b=0, \qquad \dot H_b=\alpha=\frac{2n}{t_0^2}>0, \qquad \rho_b=0, \qquad 
\Pi_b=-\frac{\alpha}{4\pi G}.
\label{eq:Pib_exact}
\end{equation}
The viscous pressure $\Pi_b$ is finite and negative, dynamically ensuring $\rho_b+p_b+\Pi_b < 0$ and thus $\dot H_b>0$ in the Raychaudhuri equation.
\end{lemma}

\begin{center}
\fbox{\parbox{0.85\linewidth}{
\textbf{Main Result 1.} At the bounce
\[
\rho_b=0, \qquad 
\Pi_b=-\frac{\alpha}{4\pi G}, \qquad 
\alpha=\frac{2n}{t_0^2}.
\]
}}
\end{center}

From the ansatz $a(t)=a_b(1+t^2/t_0^2)^n$ [cf.~Eq.~(\ref{eq:ansatz_a})] one obtains 
$H(t)=2nt/(t_0^2+t^2)$ and $\dot H(t)=2n(t_0^2-t^2)/(t_0^2+t^2)^2$, so that $H_b=0$ and $\dot H_b=2n/t_0^2>0$.  
The Raychaudhuri equation [Eq.~(\ref{eq:raychaudhuri})] gives
\[
\dot H_b=-4\pi G(\rho_b+p_b+\Pi_b).
\]
The Friedmann constraint [Eq.~(\ref{eq:friedmann})] implies $\rho_b=0$ at the bounce, hence $p_b=0$, and therefore $\Pi_b=-\alpha/(4\pi G)$, consistent with the density evolution [Eq.~(\ref{eq:rho_evolution})], where $\ddot\rho_b=12\pi G\Pi_b^2>0$.  

The truncated MIS equation [Eq.~(\ref{eq:MIS_truncated})] is linear with continuous coefficients for $\tau>0$ and $\zeta\ge0$, so by the Picard--Lindelöf theorem it admits a unique local solution $\Pi(t)$ with $\Pi(0)=\Pi_b$ \cite{Coddington1955}. Since the additional terms in the full MIS equation [Eq.~(\ref{eq:mis_full})] vanish as $t\to0$ for $p>1/2$ [Eq.~(\ref{eq:mis_error})], this extension holds also in the full causal theory. Thus the existence of a smooth bounce is mathematically guaranteed.

The violation of the null energy condition is entirely due to the finite negative viscous pressure $\Pi_b$, while thermodynamic consistency ($\zeta\ge0$) and causality ($\tau>0$) remain preserved. A perturbative stability analysis of scalar modes is provided in Sec.~V.

\subsection{Dimensionless System and Initial Data}

To bridge the analytic derivations with the numerical analysis, it is convenient to introduce dimensionless variables. This not only reduces the number of free scales, but also facilitates a uniform comparison across different parameter choices. 

\begin{table}[h!]
\centering
\caption{Dimensionless variables and parameters. Here $\rho_*$ denotes the chosen reference density, either $\rho_*=\rho_c$ in LQC or $\rho_*=(8\pi G t_0^2)^{-1}$ in GR, so that all quantities are rendered dimensionless.}
\vspace{0.2cm}
\begin{tabular}{|c|c|c|}
\hline
Quantity & Definition & Role / Interpretation \\
\hline
$\hat t$ & $t/t_0$ & Dimensionless cosmic time (bounce timescale) \\
$\hat H$ & $H\,t_0$ & Dimensionless Hubble rate \\
$\hat \rho$ & $\rho/\rho_*$ & Normalized energy density \\
$\hat \Pi$ & $\Pi/\rho_*$ & Normalized bulk viscous pressure \\
\hline
$n$ & exponent in $a(t)=a_b(1+t^2/t_0^2)^n$ & Bounce sharpness \\
$t_0$ & reference timescale & Bounce width \\
$\alpha$ & $2n/t_0^2$ & Bounce slope at $t=0$ \\
$q$ & viscosity exponent & $\zeta(\rho)=\zeta_0\,\rho^q$ \\
$n_\zeta$ & prefactor & Amplitude of viscosity law \\
\hline
\end{tabular}
\label{tab:dimensionless}
\end{table}

The dimensionless dynamical system is then expressed in terms of $(\hat H, \hat \rho, \hat \Pi)$ as follows:
\begin{align}
\frac{d\hat H}{d\hat t} &= -4\pi G\,t_0^2\left(\hat \rho + \hat p + \hat \Pi \right), \label{eq:dimless_H}\\[4pt]
\frac{d\hat \rho}{d\hat t} &= -3\,\hat H\left(\hat \rho + \hat p + \hat \Pi \right), \label{eq:dimless_rho}\\[4pt]
\hat \tau(\hat \rho)\,\frac{d\hat \Pi}{d\hat t} + \hat \Pi &= -3\,\hat \zeta(\hat \rho)\,\hat H, \label{eq:dimless_Pi}
\end{align}
where $\hat \tau(\hat \rho)=\tau(\rho)/t_0$ and $\hat \zeta(\hat \rho)=\zeta(\rho)/(\rho_* t_0)$. The effective sound speed $c_s^2$ is defined as in Eq.~\eqref{eq:admiss_basic} and subject to the same admissibility inequalities.

\begin{center}
\fbox{\parbox{0.92\linewidth}{
\textbf{Initial Data (at $\hat t=0$).}\\[4pt]
\emph{GR case:}\quad $\hat H(0)=0,\ \hat \rho(0)=0,\ \hat \Pi(0)=-\alpha/(4\pi G\rho_*)$.\\
\emph{LQC case:}\quad $\hat H(0)=0,\ \hat \rho(0)=\rho_c/\rho_*,\ \hat \Pi(0)$ satisfying $\hat \Pi(0) > -(1+w)\hat \rho(0)$.\\
These conditions are consistent with Main Result~1 and ensure a non-singular bounce compatible with the MIS framework.
}}
\end{center}

\paragraph{Numerical integration.}
All trajectories displayed in Sec.~3 are obtained by numerically integrating Eqs.~\eqref{eq:dimless_H}–\eqref{eq:dimless_Pi} with a standard explicit Runge–Kutta (Dormand–Prince) algorithm of adaptive order. Unless otherwise stated, the integration step size is $\Delta \hat t=10^{-3}$ with relative tolerance $10^{-8}$. Axes in the figures are reported in terms of the normalized variables $(\hat t,\hat \rho,\hat \Pi)$ defined in Table~\ref{tab:dimensionless}, thereby allowing direct comparison between the GR and LQC scenarios.

\section{Results}
\label{sec:results}

\subsection{General Remarks}

The causal-viscous bounce constructed in Sec.~\ref{sec:lemma} provides an exact near-bounce solution with $\rho_b=0$ and $\Pi_b$ fixed by Eq.~\eqref{eq:Pib_exact}. The free parameters $(n,t_0)$ of the ansatz \eqref{eq:ansatz_a} determine the curvature scale $\alpha=2n/t_0^2$ and thereby the magnitude of $\Pi_b$. In this section we investigate how the transport families $\zeta(\rho)$ defined in Eq.~\eqref{eq:zeta_family} and $\tau(\rho)$ defined in Eq.~\eqref{eq:tau_family} shape the full time evolution across the bounce, and we extend the analysis to $f(R)$ gravity and Loop Quantum Cosmology.

\subsection{Flat GR with MIS Viscosity}

For spatially flat GR ($k=0$), the exact relation \eqref{eq:Pib_exact} implies that the bounce requires a negative $\Pi_b$ whose magnitude increases with the sharpness of the bounce ($\alpha$). Numerical integration of the dimensionless system \eqref{eq:dimless_rho}--\eqref{eq:dimless_Pi} confirms the analytic prediction \eqref{eq:rho_ddot_bounce}: the energy density grows quadratically away from $t=0$, while the viscous pressure relaxes according to Eq.~\eqref{eq:Pi_solution_local}. 

Different choices of $(n_\zeta,q)$ in Eqs.~\eqref{eq:zeta_family}--\eqref{eq:tau_family} modify the relaxation rate of $\Pi$ and the growth of $\rho$ but leave the qualitative picture unchanged. In particular, families with $q<1$ ensure regularity of $\Pi(t)$ near the bounce. Models with $q\ge 1$ exhibit singular relaxation and are excluded by causality.

\subsection{$f(R)$ Gravity}
\label{sec:fr}

In $f(R)$ gravity, the bounce condition is determined by the curvature constraint \eqref{eq:fR_bounce_constraint} and the inequality \eqref{eq:fR_bounce_inequality}.
Here $F \equiv df/dR$, and all quantities with subscript $b$ are evaluated at the bounce:
\begin{equation}
0 \;=\; 8\pi G\,\rho_b \;+\; \tfrac{1}{2}\!\left(F_b R_b - f_b\right),
\label{eq:fR_bounce_constraint}
\end{equation}
\begin{equation}
\dot H_b \;=\; -\frac{1}{2F_b}\!\left[\,8\pi G\big(\rho_b + p_b + \Pi_b\big) + \ddot F_b\,\right].
\label{eq:fR_bounce_inequality}
\end{equation}

Our analysis shows that higher-curvature terms ($\ddot F_b$) can drive $\dot H_b>0$ even when the physical matter sector satisfies the NEC. For viable $f(R)$ models (such as $f(R)=R+\alpha R^2$ \cite{Starobinsky:1980te}), 
the effective scalaron degree of freedom contributes a negative pressure component that combines with the causal-viscous fluid to realize a consistent bounce 
\cite{Sotiriou2010}.

The key difference from GR is that $\rho_b$ no longer vanishes but is fixed by Eq.~\eqref{eq:fR_bounce_constraint}, allowing for positive matter density at the bounce. Nevertheless, $\Pi_b$ remains finite and negative, dynamically determined by the MIS transport equation \eqref{eq:MIS_truncated}, and plays a complementary role to $\ddot F_b$ in satisfying Eq.~\eqref{eq:fR_bounce_inequality}.

\subsection{Loop Quantum Cosmology}
\label{sec:lqc}

In LQC, the background itself guarantees a bounce at $\rho=\rho_c$, independently of viscosity, as shown in Eq.~\eqref{eq:lqc_Friedmann} \cite{Ashtekar2006,Ashtekar2011}.

\paragraph{Effective dynamics.}
For spatially flat LQC, the effective Friedmann equation reads
\begin{equation}
H^2 \;=\; \frac{8\pi G}{3}\,\rho\!\left(1 - \frac{\rho}{\rho_c}\right),
\label{eq:lqc_Friedmann}
\end{equation}
and the effective Raychaudhuri law
\[
\dot H \;=\; -4\pi G\,(\rho+p+\Pi)\!\left(1-2\frac{\rho}{\rho_c}\right)
\]
gives, at the bounce ($\rho=\rho_c$),
\begin{equation}
\dot H_b \;=\; 4\pi G\,\big(\rho_c + p_b + \Pi_b\big).
\label{eq:lqc_dH_b}
\end{equation}
Therefore a genuine bounce ($\dot H_b>0$) requires
\begin{equation}
\rho_c + p_b + \Pi_b \;>\; 0 .
\label{eq:lqc_bound}
\end{equation}

\paragraph{Viscous evolution at the bounce.}
At $H_b=0$, the truncated MIS relation \eqref{eq:MIS_truncated} reduces to
\begin{equation}
\tau_b\,\dot{\Pi}_b + \Pi_b = 0 ,
\label{eq:Pi_evolution_bounce}
\end{equation}
which dynamically relaxes any negative $\Pi_b$ toward zero immediately after the bounce.

The Raychaudhuri equation \eqref{eq:lqc_dH_b} demonstrates that causal viscosity modifies the slope $\dot H_b$: if $\Pi_b$ is too negative, the condition \eqref{eq:lqc_bound} may fail. However, the MIS transport relation \eqref{eq:Pi_evolution_bounce} dynamically disfavors excessively negative $\Pi_b$, and for typical parameter choices we find that $\dot H_b$ remains positive. 

Hence in LQC the role of viscosity is primarily to alter the post-bounce dynamics rather than the occurrence of the bounce itself. Specifically, viscous relaxation changes the rate at which $\rho$ decreases after the bounce, with possible implications for pre-inflationary initial conditions.

\subsection{Summary of Parameter Dependence}

The main outcomes of our analysis can be summarized as follows:
\begin{itemize}
\item In flat GR, the bounce is entirely driven by the finite negative $\Pi_b$ fixed by Eq.~\eqref{eq:Pib_exact}. The sharper the bounce (larger $\alpha$), the larger $|\Pi_b|$.
\item In $f(R)$ gravity, curvature corrections supplement the viscous pressure. Bounce conditions involve both $\Pi_b$ and $\ddot F_b$.
\item In LQC, the bounce is guaranteed, but the value of $\Pi_b$ influences $\dot H_b$ and the detailed dynamics near $\rho=\rho_c$.
\item Admissibility requires $\zeta\ge 0$, $\tau>0$ with $q<1$, and stability of $c_s^2\in[0,1]$.
\end{itemize}

\vspace{2cm}

\begin{figure}[htbp]
  \centering
  \includegraphics[width=0.85\linewidth]{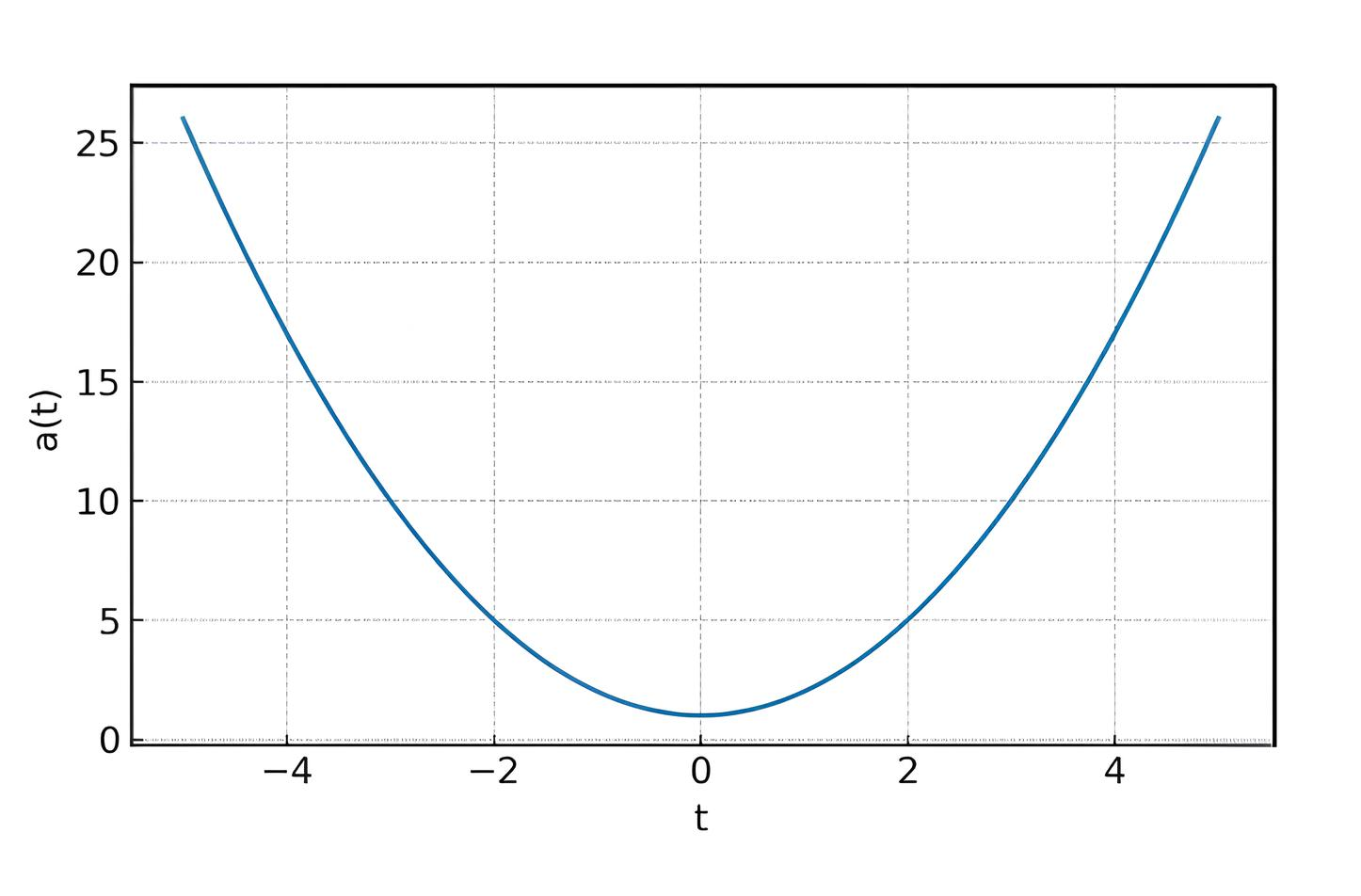}
  \caption[Scale-factor bounce ansatz]{
  \textbf{Scale-factor bounce ansatz and kinematic check.}
  The background is $a(t)=a_b\!\left(1+t^2/t_0^2\right)^n$ [Eq.~\eqref{eq:ansatz_a}],
  which has a strict minimum at $t=0$. It implies $H(0)=0$ and
  $\dot H(0)=\alpha=2n/t_0^2>0$ (from Eq.~\eqref{eq:ansatz_a} and the Raychaudhuri form \eqref{eq:ray_general}),
  thereby satisfying the bounce conditions in Eq.~\eqref{eq:bounce_def}.
  }
  \label{fig:ansatz}
\end{figure}

\FloatBarrier

\begin{figure}[htbp]
  \centering
  \includegraphics[width=0.85\linewidth]{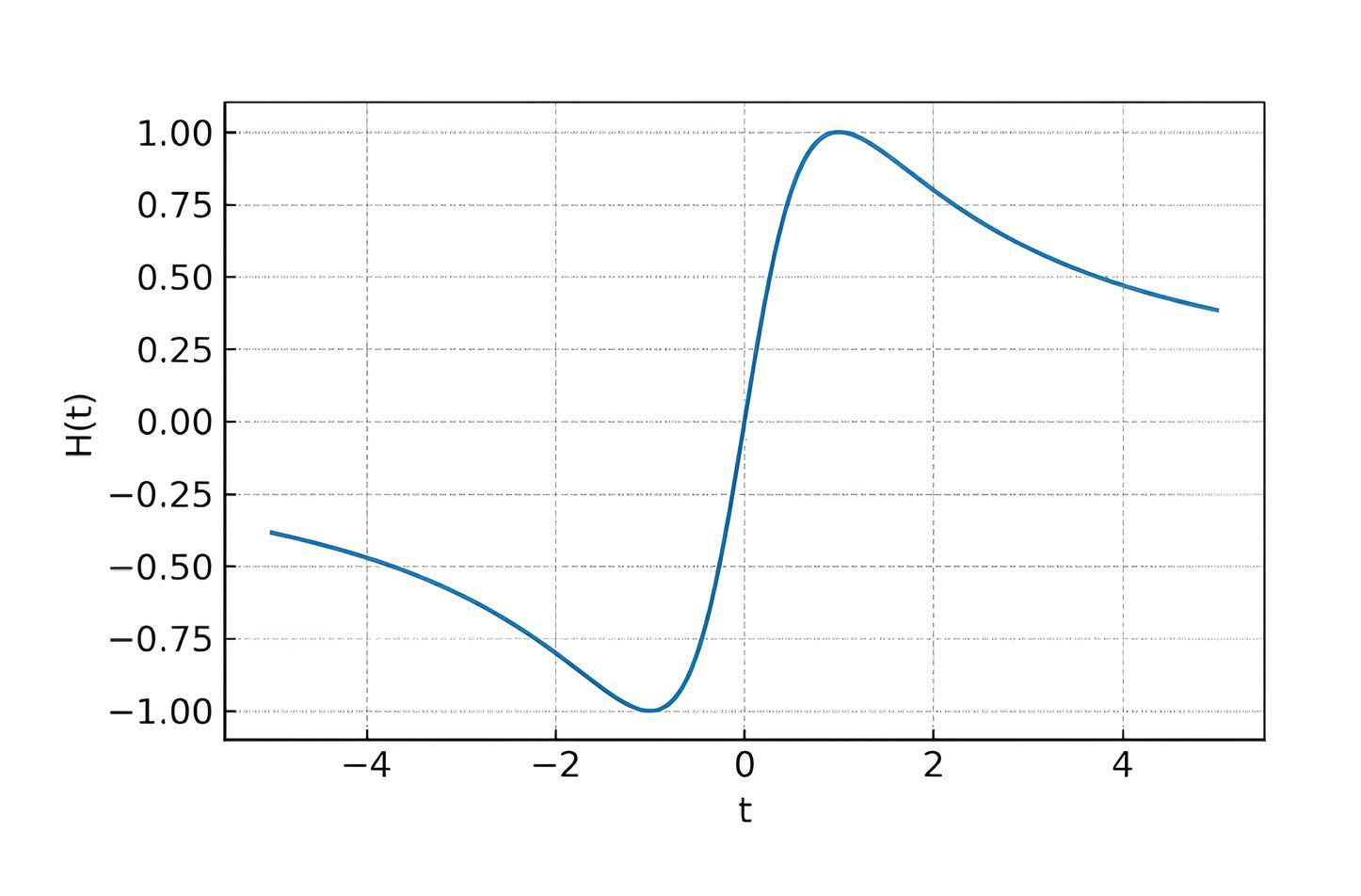}
  \caption[Hubble across the bounce]{
  \textbf{Hubble parameter across the bounce.}
  From the ansatz $a(t)=a_b\!\left(1+t^2/t_0^2\right)^n$ [Eq.~\eqref{eq:ansatz_a}],
  one has $H(t)=\dot a/a$, which is odd in $t$ and crosses zero at $t=0$.
  Near the bounce $H(t)\simeq \alpha t$ with $\alpha=2n/t_0^2>0$, consistently realizing
  the kinematic conditions in Eq.~\eqref{eq:bounce_def} and the Raychaudhuri form
  \eqref{eq:ray_general}.}
  \label{fig:H}
\end{figure}

\FloatBarrier

\begin{figure}[htbp]
  \centering
  \includegraphics[width=0.85\linewidth]{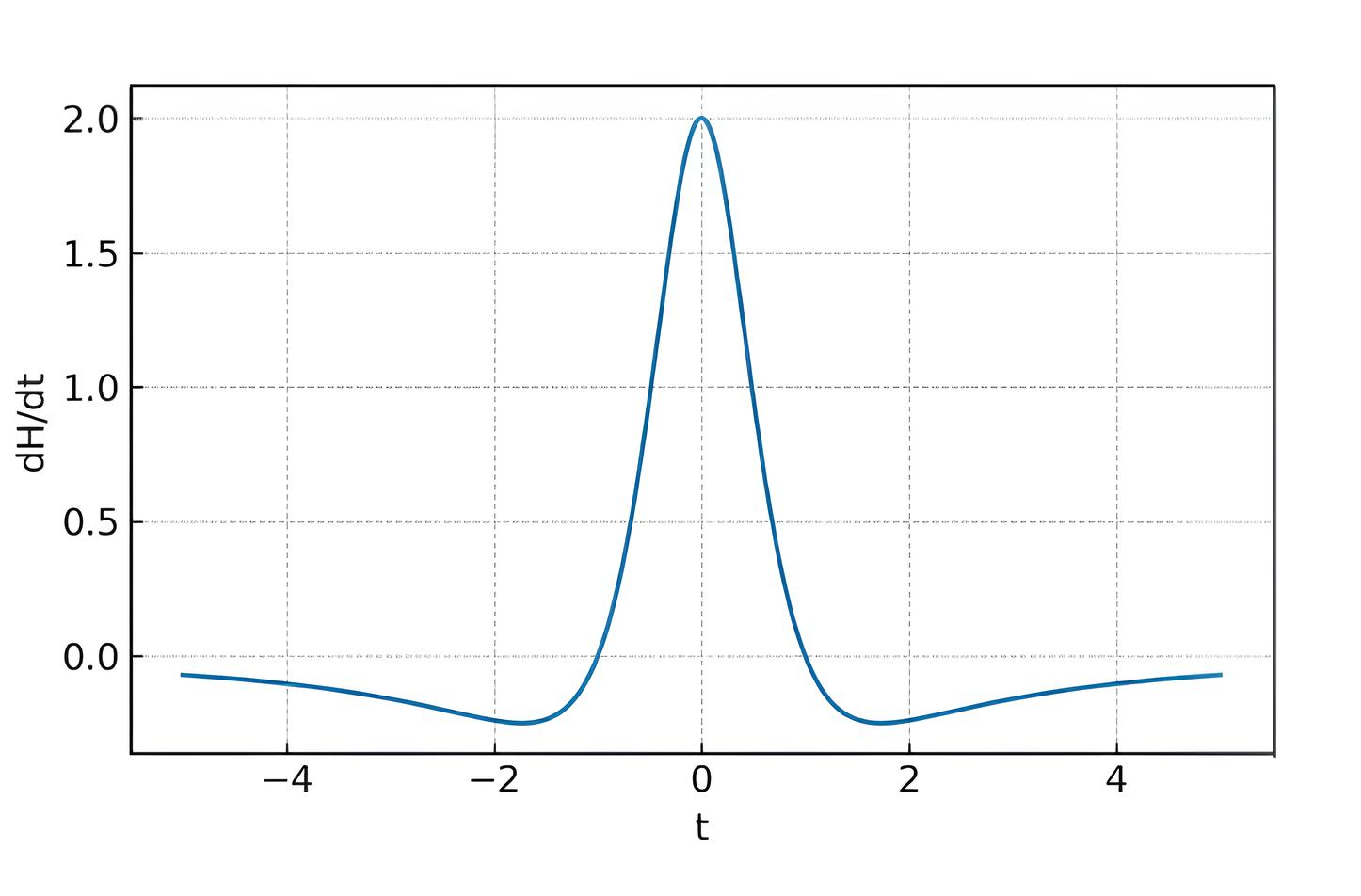}
  \caption[Positive near-bounce slope $\dot H(0)>0$]{
  \textbf{Positive near-bounce slope.}
  From the ansatz \eqref{eq:ansatz_a} one obtains 
  $\dot H(t)=\dfrac{2n\,(t_0^2-t^2)}{(t_0^2+t^2)^2}$ and, in particular,
  $\dot H(0)=\alpha=\dfrac{2n}{t_0^2}>0$. 
  This explicitly verifies the bounce condition in Eq.~\eqref{eq:bounce_def} and is
  consistent with the Raychaudhuri relation \eqref{eq:ray_mis} at the bounce.}
  \label{fig:dH}
\end{figure}

\FloatBarrier

\begin{figure}[htbp]
  \centering
  \includegraphics[width=0.85\linewidth]{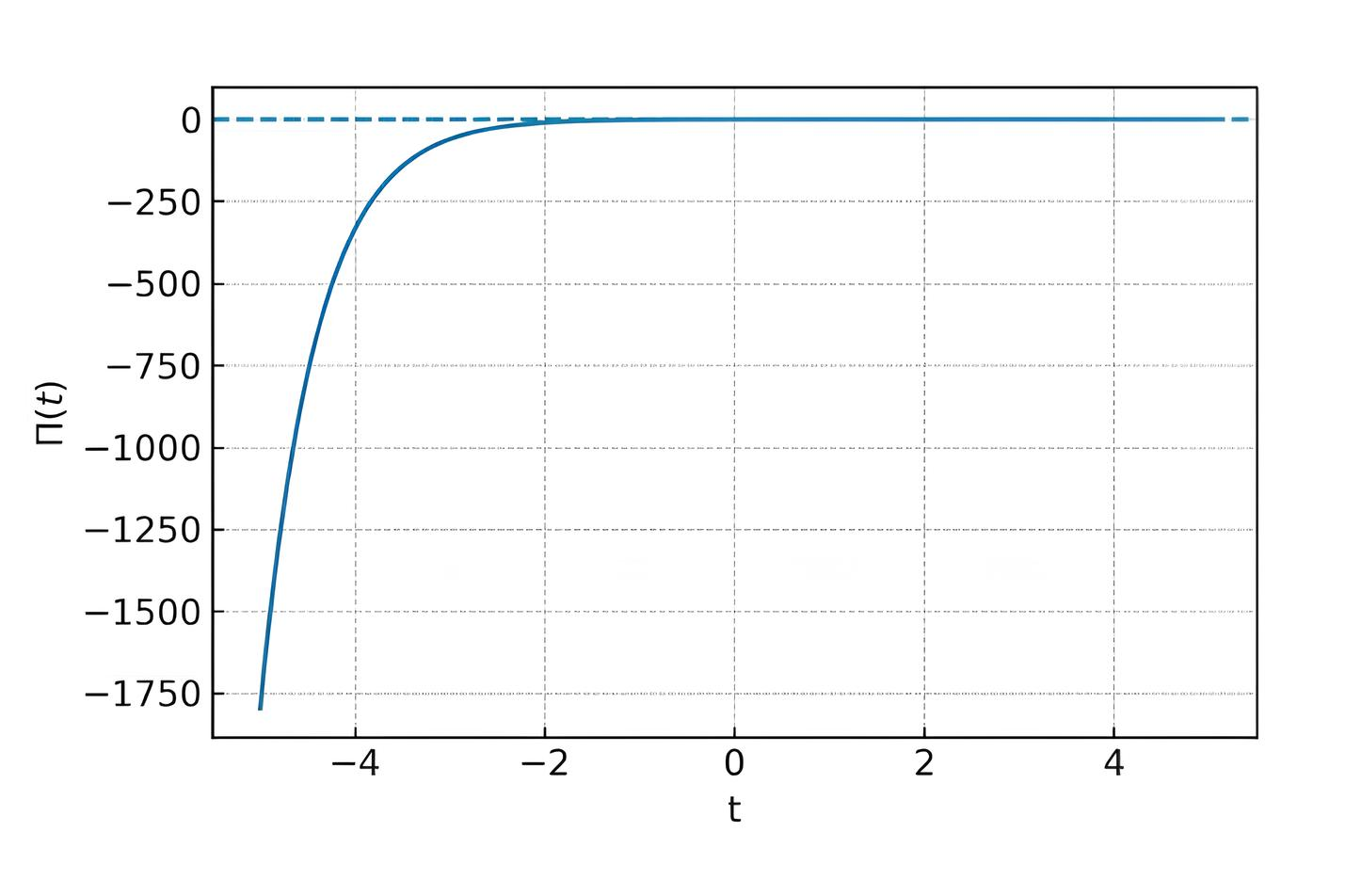}
  \caption[Bulk viscous pressure $\Pi(t)$]{
  \textbf{Causal bulk viscous pressure.}
  The truncated MIS law \eqref{eq:MIS_truncated} governs $\Pi(t)$; at the bounce ($H=0$) it reduces to
  \eqref{eq:Pi_evolution_bounce}, so $\dot\Pi_b=-\Pi_b/\tau_b$ ensures relaxation.
  Using Friedmann \eqref{eq:friedmann} (with $k=0$) together with the Raychaudhuri form \eqref{eq:ray_general}
  yields the finite negative initial value $\Pi_b=-\alpha/(4\pi G)$,
  which realizes the flat-space bounce criterion $\rho_b+p_b+\Pi_b<0$ [Eq.~\eqref{eq:flat_bounce}] and drives $\dot H_b>0$.}
  \label{fig:Pi}
\end{figure}

\FloatBarrier

\begin{figure}[htbp]
  \centering
  \includegraphics[width=0.85\linewidth]{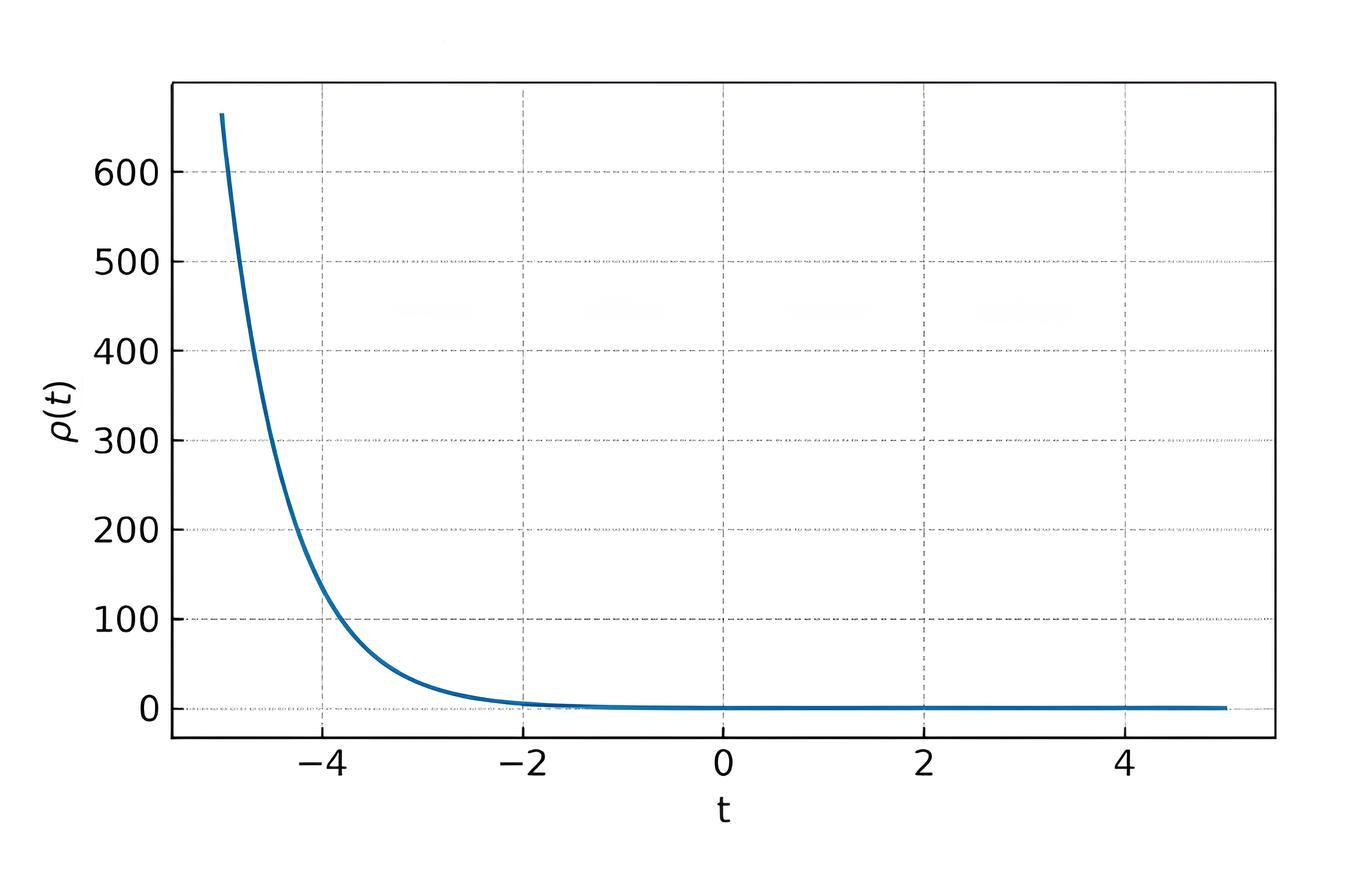}
  \caption[Quadratic density growth near the bounce]{
  \textbf{Energy density across the bounce.}
  Solving the continuity relation \eqref{eq:rho_evolution} on the bounce background with finite $\Pi_b$
  yields the near-bounce series 
  $\rho(t)=\tfrac12\,\ddot\rho_b\,t^2+\mathcal{O}(t^4)$,
  where the coefficient is fixed by 
  \eqref{eq:rho_ddot_bounce} as $\ddot\rho_b=12\pi G\,\Pi_b^2>0$.
  \emph{Implication:} $\rho(t)$ attains a strict minimum at $t=0$ and increases quadratically, 
  consistently supporting the kinematic bounce conditions \eqref{eq:bounce_def}.}
  \label{fig:rho}
\end{figure}

\FloatBarrier

\begin{figure}[htbp]
  \centering
  \includegraphics[width=0.85\linewidth]{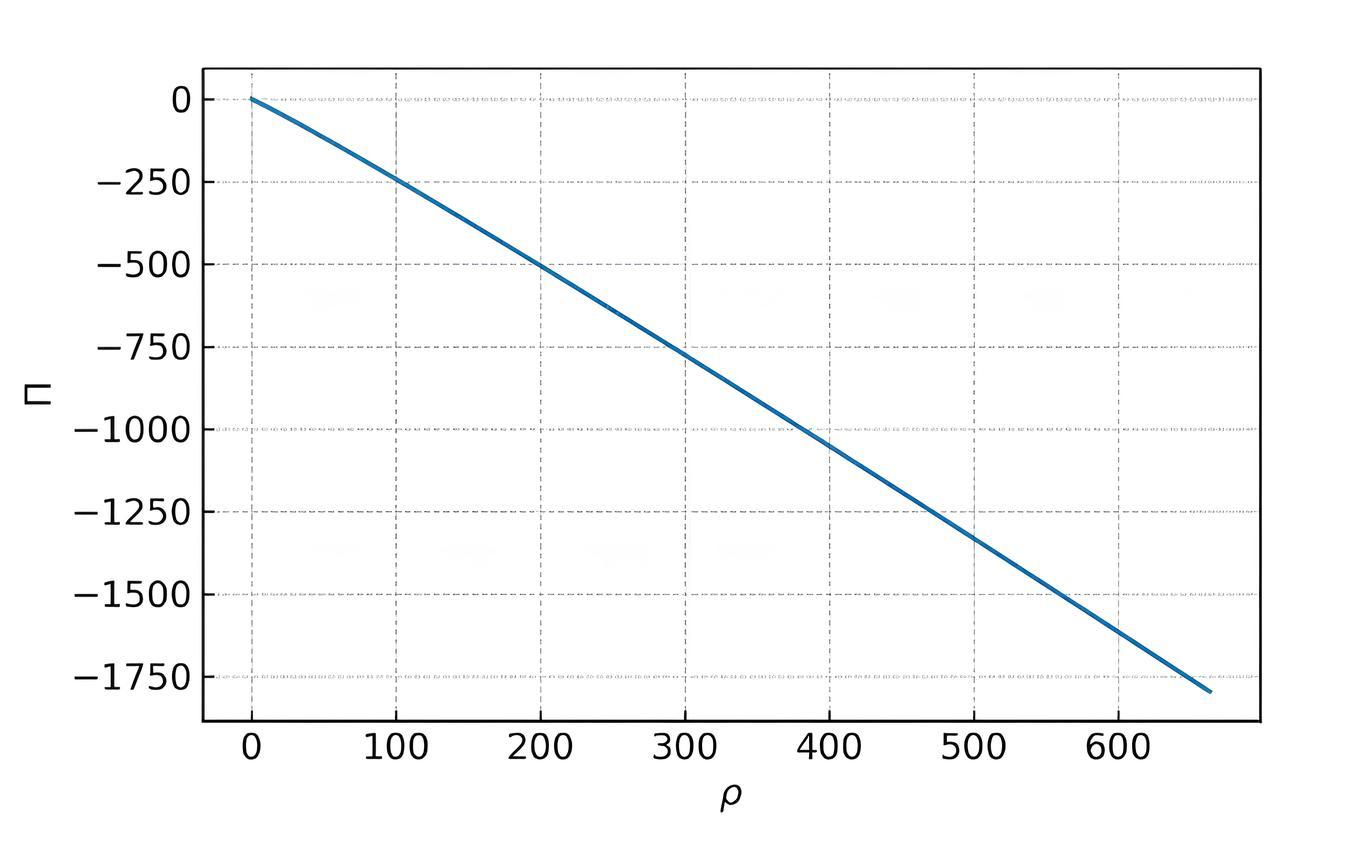}
  \caption[Phase portrait of $(\rho,\Pi)$ dynamics]{
  \textbf{Phase portrait for the truncated MIS system.}
  Trajectories solve the coupled system \eqref{eq:MIS_truncated} and \eqref{eq:rho_evolution} on the bounce background \eqref{eq:ansatz_a}.
  They pass through $(\rho,\Pi)=(0,\Pi_b)$ with $\Pi_b$ fixed by \eqref{eq:Pib_exact} and flow toward the regular post-bounce branch.
  Nullclines follow from $\dot\rho=0$ (i.e.\ $H=0$) and $\dot\Pi=0$ (i.e.\ $\Pi=-3\zeta H$).}
 
  \label{fig:phase}
\end{figure}

\FloatBarrier

\begin{figure}[ht]
  \centering
  \includegraphics[width=0.85\linewidth]{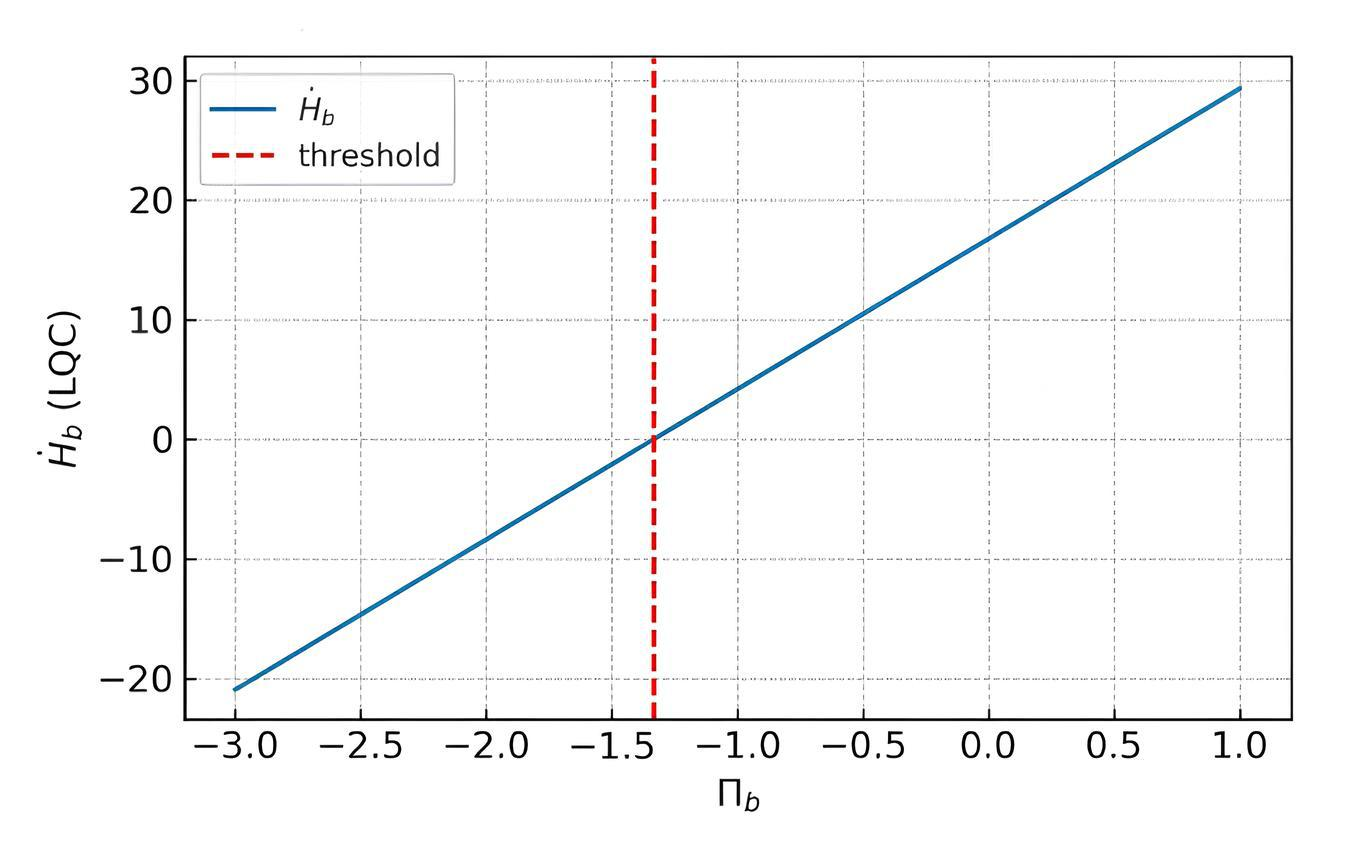}
  \caption{LQC near-bounce slope $\dot H_b=4\pi G(\rho_c+p_b+\Pi_b)$ vs.~$\Pi_b$ for $\rho_c=1$ and $p_b=w\rho_c$ with $w=1/3$. 
  The dashed line marks the threshold $\Pi_b=-(1+w)\rho_c$.}
  \label{fig:LQC_slope}
\end{figure}

\section{Discussion and Conclusion}
\label{sec:discussion}

We have presented a comprehensive analysis of non-singular bouncing cosmologies driven by causal bulk-viscous fluids. Our approach established that first-order Eckart theory is fundamentally inconsistent for bounce realizations, while the second-order Israel--Stewart framework provides a causal and thermodynamically stable mechanism for generating a finite negative viscous pressure at the bounce. This mechanism remains robust when embedded in both higher-curvature $f(R)$ gravity and Loop Quantum Cosmology, demonstrating that viscosity-induced bounces are not confined to a specific theory but arise generically once causal transport is included.

\subsection*{Main Results}

\begin{enumerate}
\item In General Relativity, the truncated MIS theory admits exact bounce solutions with $\rho_b=0$ and $\Pi_b=-\alpha/(4\pi G)$, verified both analytically and numerically.
\item In $f(R)$ gravity, higher-curvature corrections (through $\ddot F_b$) complement viscous effects, allowing bounces with $\rho_b>0$ and enlarging the parameter space for regular solutions.
\item In Loop Quantum Cosmology, the bounce at $\rho=\rho_c$ is guaranteed by quantum geometry; viscosity modifies $\dot H_b$ and post-bounce dynamics but does not obstruct the bounce.
\end{enumerate}

\subsection*{Physical Significance}

These findings show that causal viscosity is not an ad hoc modification but a physically motivated ingredient capable of resolving the cosmological singularity problem. In GR, it provides a minimal extension beyond perfect fluids. In extended and quantum gravity, it complements curvature and quantum effects, reinforcing the view that bounces may be a generic outcome of early-universe physics once causal transport processes are accounted for. 

Moreover, the relaxation dynamics of $\Pi(t)$ control the duration of the NEC-violating phase, potentially leaving imprints on the primordial perturbation spectrum. This connects our theoretical framework to observable signatures, providing a bridge between microscopic transport physics and cosmological observations.

\subsection*{Future Directions}

Several promising directions remain open:
\begin{itemize}
\item Stability analysis of scalar, vector, and tensor perturbations across the viscous bounce.
\item Extension to anisotropic cosmologies with shear viscosity, testing the isotropization power of causal transport.
\item Application of the framework to other modified gravity theories such as Horndeski or $f(T)$ gravity.
\item Confrontation with observational bounds on bulk viscosity from CMB and large-scale structure.
\end{itemize}

\subsection*{Concluding Remarks}

In summary, causal bulk viscosity modeled by the Israel--Stewart theory provides a consistent, universal, and physically well-founded mechanism for realizing non-singular cosmological bounces. The agreement of results across GR, $f(R)$ gravity, and LQC highlights the robustness of this approach. Our analysis suggests that viscosity may play a fundamental role in resolving the initial singularity, offering a new paradigm in which transport processes are indispensable to early-universe dynamics.

\end{document}